\documentclass[journal=jpclcd,manuscript=article]{achemso}
\usepackage{color}
\usepackage{graphicx}
\usepackage{caption}

\usepackage[version=3]{mhchem} % Formula subscripts using \ce{}

\author{Jiuyu Sun}
\author{Ruiqi Zhang}
\author{Xingxing Li}
\affiliation[Hefei National Laboratory for Physical Sciences at the Microscale, University of Science and Technology of China]
{Hefei National Laboratory for Physical Sciences at the Microscale, University of Science and Technology of China, Hefei, Anhui 230026, China}
\alsoaffiliation[ICQD, University of Science and Technology of China]
{International Center for Quantum Design of Functional Materials (ICQD), University of Science and Technology of China, Hefei, Anhui 230026, China}
\author{Jinlong Yang}
\email{jlyang@ustc.edu,cn}
\phone{+86-551-63606408}
\fax{ +86-551-63603748 (J. Y.)}
\affiliation[Hefei National Laboratory for Physical Sciences at the Microscale, University of Science and Technology of China]
{Hefei National Laboratory for Physical Sciences at the Microscale, University of Science and Technology of China, Hefei, Anhui 230026, China}
\alsoaffiliation[ICQD, University of Science and Technology of China]
{International Center for Quantum Design of Functional Materials (ICQD), University of Science and Technology of China, Hefei, Anhui 230026, China}

\title[An \textsf{achemso} demo]
{A many-body \textit{GW}+BSE investigation of electronic and optical properties of C$_{2}$N}

\begin{document}


%\begin{abstract}

ABSTRACT: A newly synthesized layered material C$_{2}$N was investigated based on many-body perturbation theory using \textit{GW} plus Bethe-Salpeter equation approach. The electronic band gap was determined to be ranging from 3.75 to 1.89 eV from monolayer to bulk. Significant GW quasiparticle corrections, of more than 0.9 eV, to the Kohn-Sham band gaps from the local density approximation (LDA) calculations are found. Strong \textit{excitonic effects} play a crucial role in optical properties. We found large binding energies of greater than 0.6 eV for bound excitons in few-layer C2N, while it is only 0.04 eV in bulk C$_{2}$N.  All the structures exhibit strong and broad optical absorption in the visible light region, which makes C$_{2}$N a promising candidate for solar energy conversion, such as photocatalytic water splitting.

%\end{abstract}

\vspace{3ex}



\section{Introduction}

Since the dicovery of graphene\cite{graphene-first},  two-dimensional
(2D) and quasi-2D materials are of tremendous interest predominantly
due to their extraordinary physical or chemical properties and great potential of application\cite{graphene-GW-prl,graphyne-prl,MoS2-nanolett,graphene-BN}. Graphitic C$_{3}$N$_{4}$ (g-C$_{3}$N$_{4}$) has attracted much attention in recent years, for its high stability and
moderate band gap, which can function  as a
metal-free polymeric photocatalyst for splitting water molecules
with solar energy\cite{c3n4-nature}. A new carbon nitride C$\_2$N was also successfully synthesized most recently, which shows an optical gap of 1.96 eV and high on/off ratio of 10$^{7}$\cite{Expt}. The C$_{2}$N
crystal is a layered structure with
uniform holes and nitrogen atoms, as called 'C$_{2}$N-\textit{h}2D', which is composed of atomically thin C2N monolayers [Fig.~\ref{fgr:mono}(a)] stacked by Van Der Waals interactions. Follow-up works suggest C$_{2}$N to be a promising metal-free photocatalyst for splitting water molecules with visible light by density functional theory (DFT) calculations\cite{Ruiqi-nl,Ruiqi-C2N-splitting} and a hydrogen evolution experiment\cite{c2n_catalyst_expt}. 

However, one cannot realistically envision C$_{2}$N devices until its fundamental excited-state properties, such as quasiparticle (QP) band and optical spectrum, are obtained. In 2D materials, Coulomb interactions between excited electrons and holes , i.e.,
\textit{excitonic effects}, play a very important role in optical
properties\cite{graphene-BSE,graphene-BSE-prl,PhysRevLett.104.226804,BN-BSE-prl,MoS2-BSE-prl,BP-prb,Germanane}.
Quantum size confinement and less efficient electronic screening are
two well-defined factors contributing to the excitonic
effects\cite{graphene_ribbon_nanolett}. 
Owing to its layered structure, C$_{2}$N-\textit{h}2D is expected to have strong excitonic effects
too. Therefore, an appropriate description of the \textit{e-e}
correlation and \textit{e-h} interaction is highly required to investigate the electronic and optical properties of
C$_{2}$N-\textit{h}2D.

In the paper, we perform first-principles \textit{GW} plus Bethe-Salpeter equation (BSE)\cite{BSE} simulations to study the QP band structures and optical spectra of few-layer and bulk  C$_{2}$N-\textit{h}2D. Firstly, the QP band gap of monolayer C$_{2}$N enlarges from 1.7 eV (obtained by normal LDA calculations) to 3.75 eV, and due to strong excitonic effects, the lowest absorption peak is reduced to 2.75 eV compared to the electronic band gap. Secondly, the electronic band gaps, exciton binding energies, first bright excitonic energies (first optical absorption peaks) reduce with the increasing of the number of layers. Finally, strong optical absorptions in the visible light region were found for all structures from monolayer to bulk C$_{2}$N-\textit{h}2D. Our results suggest that few-layer C$_{2}$N-\textit{h}2D has better visible light  absorption than graphitic C$_{3}$N$_{4}$ (g-C$_{3}$N$_{4}$)\cite{c3n4-prb},  which makes C$_{2}$N a promising candidate for metal free photocatalyst with enhanced solar energy conversion efficiency.

\section{Methods}
In order to provide a good staring point for many-body calculations
within the \textit{GW} approximation, all the ground state of
C$_{2}$N structures were calculated using density functional theory
(DFT) with local density approximation (LDA) functional as implemented in the  QUANTUM ESPRESSO
code\cite{QE}. The plane-wave cutoff was set as 50 RY with a
norm-conserving pseudopotential. A $10\times10\times1$ and $8\times8\times6$ \textbf{k} grid was used for few-layered structures and bulk, respectively. The structures were fully relaxed until an energy convergence of $10^{-9}$ RY and a force convergence on atoms of 0.01 eV/\r{A}.

Starting from the wavefunction and energies of Kohn-Sham equations,
the quasiparticle \textit{GW} and BSE calculations were
performed by using BerkeleyGW package\cite{BGW}. A dynamical dielectric matrix within the random phase approximation (RPA) scheme and a generalized plasmon-pole model\cite{GPP}. This gives rise to QP energy within a single-shot $G_{0}W_{0}$ calculation. The involved unoccupied band number is up to more than 5 times of occupied band for every structure to achieve the converged
dielectric function. The excitonic effects (\textit{e-h} interactions) are  included by solving BSE with $30\times30\times1$ and  $16\times16\times12$ \textbf{k} grids for few-layered structures and bulk, respectively.

\section{Results and discussion}

\subsection{QP band structure}

\begin{figure}
\centering
\includegraphics[width=8.5cm]{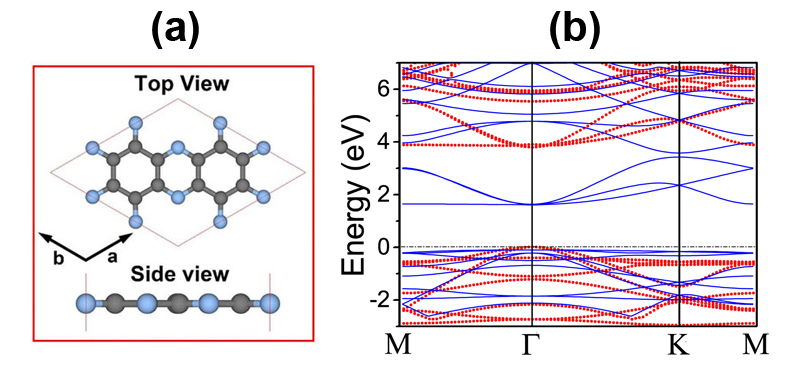}
\caption{The (a) geometry structure and (b) band structure calculated by LDA (blue line) and \textit{GW} (red dot) of monolayer C$_{2}$N-\textit{h}2D. The grey and blue balls represent C atoms and N atoms, respectively.}
\label{fgr:mono}
\end{figure}

%\begin{figure}
%\begin{center}
%\includegraphics[width=8.5 cm]{figure1.jpg}
%\end{center}
%\caption{The (a) geometry structure and (b) band structure calculated by LDA (blue line) and \textit{GW} (red dot) of monolayer C$_{2}$N-\textit{h}2D. The grey and blue balls represent C atoms and N atoms, respectively.}
%\label{fgr:mono}
%\end{figure}

The equivalent lattice parameter was optimized to 8.28
\r{A} for monolayer C$_{2}$N-\textit{h}2D, which is consistent to 8.24$\pm$0.96 \r{A} by the
experiment\cite{Expt}. %\subsection{Band structures}
All of the band gap results calculated by DFT-LDA and \textit{GW}
method are shown in Table~\ref{tbl:gaps}, as well as the results by HSE06\cite{Ruiqi-C2N-splitting}.

\begin{table}
  \caption{The band gap energies $E_{g}$ calculated by LDA, HSE06 and \textit{GW} of monolayer, bilayer and bulk C$_{2}$N-\textit{h}2D. For bulk, we listed $E_{g}$ for both direct and indirect gaps. $\Delta_{1}$ and $\Delta_{2}$ were defined as: $\Delta_{1}=E_{g}^{HSE06}-E_{g}^{LDA}$ and $\Delta_{2}=E_{g}^{GW}$-$E_{g}^{LDA}$, respectively. All values are in eV.}
  \label{tbl:gaps}
%  \begin{center}
  \begin{tabular}{lcclcc}
  \hline
          &   LDA & HSE06& \textit{GW} & $\Delta_{1}$  & $\Delta_{2}$ \\ \hline
 Monolayer   &1.71& 2.47  &  3.75   &  0.76 &  2.04  \\
 Bilayer  & 1.40  & 2.17  & 3.03 &0.77 & 1.63  \\
Trilayer  & 1.24 & 1.97 & 2.77 & 0.73 &  1.53 \\
 Bulk   & 1.11 & 1.81  & 2.07 & 0.70  & 0.96 \\
           &   1.06\textsuperscript{\emph{a}}  & 1.66\textsuperscript{\emph{a}}  &  1.89\textsuperscript{\emph{a}}  &  0.60 & 0.94  \\\hline
  \end{tabular}
%  \end{center}

  \textsuperscript{\emph{a}} Indirect band gap
\end{table}

For monolayer C$_{2}$N-\textit{h}2D, the direct band gap is located
at $\Gamma$ point. The band structures are shown in Figure~\ref{fgr:mono}(b). 
The energy gap calculated by \textit{GW} is 3.75 eV, which is much greater than that by
DFT-LDA (1.71 eV). Such large QP corrections to LDA eigenvalues are
attributed to the enhance of \textit{e-e} interactions. This
nonlocal behavior can not be correctly described by LDA. However,
HSE06 functional, having considered nonlocal exchange effects, results in a remarkably smaller gap than \textit{GW} gap. Therefore the coulomb
screening effect, which is only well described by self-energy
operator in the \textit{GW} scheme, plays a significant role in the
electronic structure of  monolayer C$_{2}$N-\textit{h}2D. It is also noted that monolayer C$_{2}$N-\textit{h}2D has comparable delocalized valence band maximum (VBM) and conduction band minimum (CBM) to g-C$_{3}$N$_{4}$\cite{c3n4-prb}. With a smaller direct gap, C$_{2}$N-\textit{h}2D is suggested to have a better performance than g-C$_{3}$N$_{4}$ in light catalyst.

\begin{figure}
\centering
\includegraphics[width=17cm]{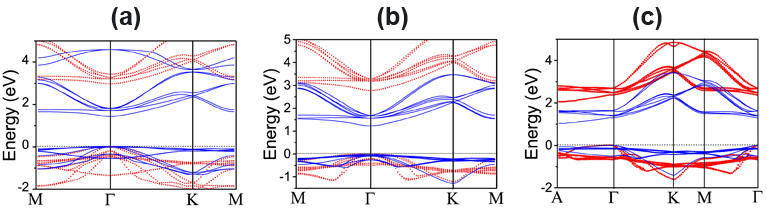}
\caption{The band structures of (a)bilayer, (b)trilayer and (c)bulk C$_{2}$N-\textit{h}2D. The band structures by LDA and \textit{GW} are plotted by blue line and red dot, respectively.}
\label{fgr:bands}
\end{figure}

Regarding to bilayer, trilayer and bulk structures, we only considered the most stable stacking mode of AB-, ABC-  and ABC-stakcing\cite{Ruiqi-nl}, respectively. All three band structures are performed in Figure~\ref{fgr:bands}. Direct band gaps at $\Gamma$ point are found in bilayer and trilayer. Compared to monolayer, the QP band gaps decrease to 3.03 eV for bilayer and 2.77 eV for trilayer. This is because of the increasing coulomb screening effect, which is induced by quantum size confinement. However, it becomes  an indirect gap for bulk (1.89 eV). The VBM occurs at $\Gamma$ point. The CBM is located at A point (0.0  0.0 0.5) in Brillouin zone, which is 0.32 eV lower than CBM at $\Gamma$. The minimum direct band gap (2.21 eV) was located near (0 0 1/3) point.

For all the band structures of C$_{2}$N-\textit{h}2D, there is no scissor operation for \textit{GW} corrections in all the presented band strutures. After considering nolocal exchange interaction, several arched valence bands moves upward across the flat VBM in LDA, while the structures of conduction bands remain unchanged. The exchange correction induced change of VBM significantly enhances the carrier mobility for C$_{2}$N-\textit{h}2D.  Then we define the HSE06 and \textit{GW} corrections to LDA band gap as $\Delta_{1}$ and $\Delta_{2}$, respectively. As listed in Table~\ref{tbl:gaps}, $\Delta_{1}$ maintains about 0.7 eV for all the few-layered structures calculated. In contrast, as the number of layer increases,  $\Delta_{2}$ decreases rapidly from 2.04 eV (monolayer) to 1.53 eV (trilayer). This indicates that the correction of coulomb screening effects, unlike \textit{e-e} exchange effects,  depend deeply in the dimensional of the system. Thus, \textit{e-e} interactions should be carefully treated when studying the electronic structure of few-layered C$_{2}$N-\textit{h}2D.  A power law fitting of the form ($A/N^{\alpha} + B$) is applied for \textit{GW} calculated gaps, where N is the number of layers. It shows that the evolution of QP band gaps follows a $1/N^{0.69}$ power law.

\subsection{Excitons and optical absorptions}

\begin{figure}
\centering
\includegraphics[width=8.5cm]{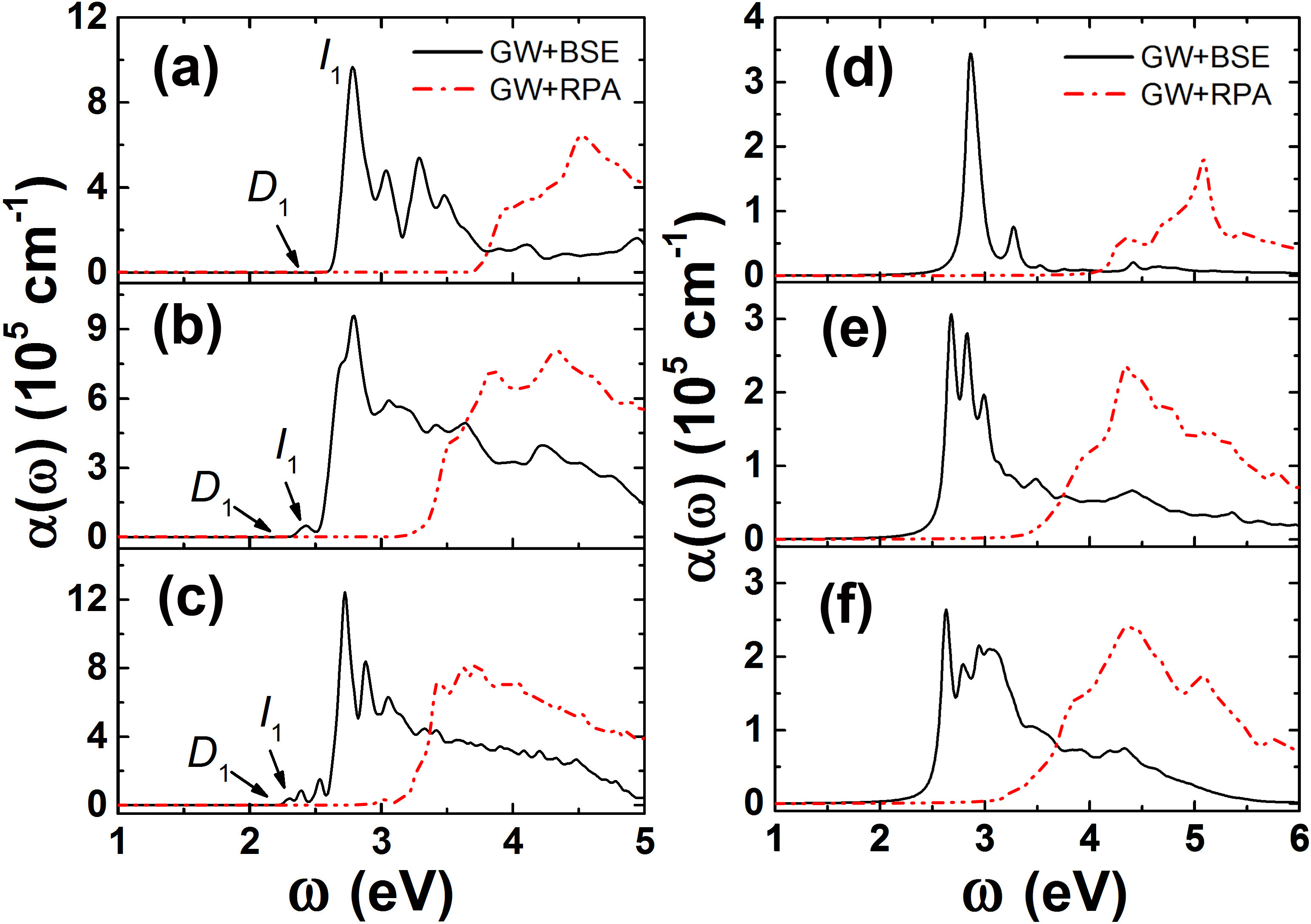}
\caption{The absorption coefficient by \textit{GW}+BSE (black line) and \textit{GW}+RPA (red dash-dot) of (a, d) monolayer, (b, e) bilayer and (c, f) tirlayer C$_{2}$N-\textit{h}2D for light polarization parallel (left side)  and perpendicular (right side) to the surface plane.  A Gaussian broadening of 0.05 eV is adopted.}
\label{fgr:abs}
\end{figure}

In Figure~\ref{fgr:abs}(a) to (f), we performed optical absorption spectrum of few-layered C$_{2}$N-\textit{h}2D. As shown in the figures, dramatic changes occur after \textit{e-h} interactions are included (\textit{GW}+BSE). The optical absorptions are greatly modified and strongly bound exciton states well below the onset of single-particle transition continuum are indicated. There is an obvious cancellation effect between the band-gap opening due to the QP corrections and the redshift of optical absorption due to the strong excitonic effects.

\begin{table}
\caption{First bright excitonic energies (optical gaps) $E_{I^{1}}$, second excitonic energies $E_{I^{2}}$, first dark excitonic energies $E_{D}$ and binding energy of first bright excitons $E^{b}_{I^{1}}$ of  monolayer, bilayer, trilayer and bulk C$_{2}$N-\textit{h}2D. All values are in eV. }
%\begin{center}
\begin{tabular}{llcccc}
\hline

     &       $E_{I^{1}}$     & $E_{I^{2}}$   & $E_{D}$ &   $E^{b}_{I^{1}}$\\ \hline
Monolayer & 2.75     & 2.76 & 2.53   &  1.11     \\
 Bilayer    & 2.42   & 2.42 & 2.36 &  0.76\\
  Trilayer &  2.30  & 2.39 & 2.26 & 0.53  \\
     Bulk   & 2.02   & 2.17 & -  & 0.04\\\hline
\end{tabular}
%\end{center}
\end{table}

In the case of monolayer,  the first optically active (bright) exciton energies at 2.75 eV ($I^{1}$) and the second one ($I^{2}$) at 3.02 eV for light parallel to surface plane. $I^{1}$ is in the nature of transitions from VBM to CBM at the $\Gamma$ point. The binding energy $E_{b}$ of the first exciton $I^{1}$ is 1.11 eV. Such a large binding energy is due to the less efficient screening in monolayer structure.  There are many optically inactive (dark) excitons.  The first dark exciton was located at 2.53 eV ($D^{1}$), dipole forbidden ($10^{-6}$). These dark excitons may affect the luminescence yield in certain cases.

The optical absorption spectra of bilayer and trilayer of C$_{2}$N-\textit{h}2D were performed in Figure~\ref{fgr:abs}(b) and (c), for light polarization parallel to the surface plane. Excitonic effects become weaker as the number of layers increasing. Comparing to monolayer, the red shifted of absorption of \textit{GW}+BSE against \textit{GW}+RPA is smaller than that in both cases of bilayer and trilayer C$_{2}$N-\textit{h}2D. This indicates smaller binding energies of $I^{1}$. The absorption peaks of first light excitons of both structures are no longer as sharp as that in monolayer. With the increasing number of layers, intrinsic absorption begins to play a dominant role, instead of excitonic absorptions. For bilayer, the first two (doubly degenerated) optical bright exciton ($I^{1}$/$I^{2}$) was located at 2.42 eV, with a binding energy of 0.76 eV. However, unlike monolayer and bilayer, $I^{1}$ and $I^{2}$ of trilayer are located at 2.30 eV and 2.39 eV, which exhibit two separate peaks in the spectra. As expected, the binding energy of $I^{1}$ in trilayer reduced to 0.53 eV. Both $I^{1}$ of bilayer and trilayer are excited from VBM to CBM near $\Gamma$ point. In addition, the energies of two $D^{1}$ are 2.36 eV and 2.26 eV eV for bilayer and trilayer, respectively. The dark excitons become stronger with dipoles in the magnitude of $10^{0}$, though there is no obvious corresponding peak in the absorption spectra. 

It draws our highly attention that there are also strong excitonic effects in few-layered structures for light polarization perpendicular to the layer plane (see Fig.~\ref{fgr:abs}(d) to (e)). The first peak of monolayer was observed only a slight blue shift (0.1 eV) against the case for parallel polarized light. This is quite different to common 2D materials (see in supporting information), whose absorptions begin with the onset of single-particle transition (electronic band gap) for  perpendicular polarized light. As shown in the figures, there are obvious red shifts compared to absorptions without \textit{e-h} interactions for all the few-layered systems, indicating huge excitonic effects along the corresponding directtion. Such a property makes few-layered C$_{2}$N-\textit{h}2D exhibits considerable absorptions of visible light along all the directions, which enhances the productivity of sunlight catalysis.

\begin{figure}
\centering
\includegraphics[width=85mm]{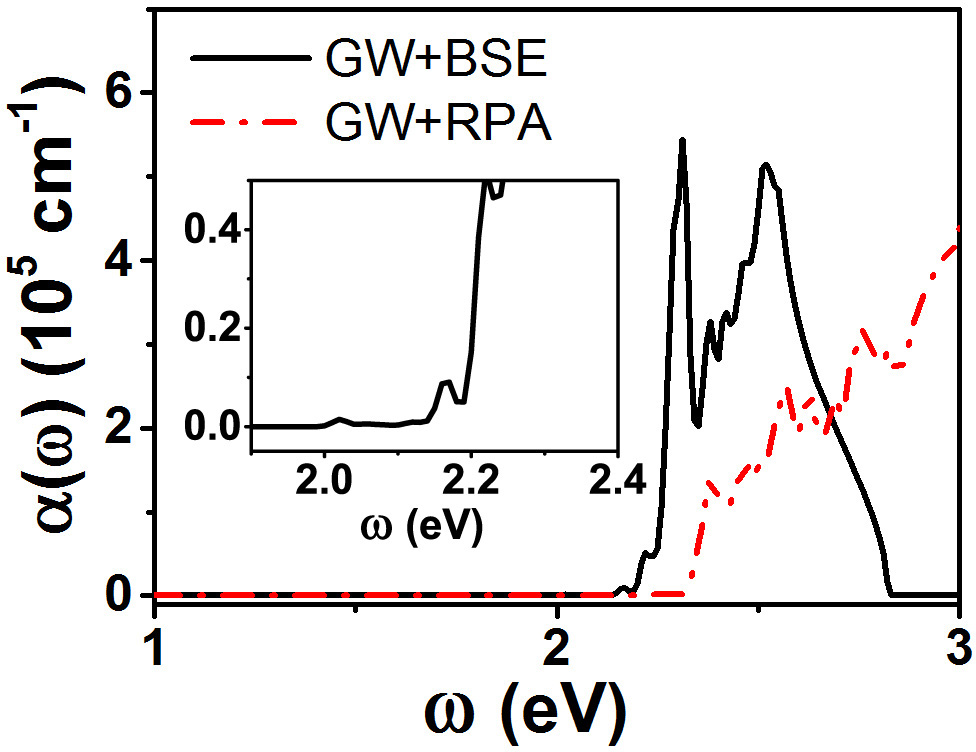}
\caption{The absorption coefficient by \textit{GW}+BSE (black line) and \textit{GW}+RPA (red dash-dot) of bulk C$_{2}$N-\textit{h}2D for light polarization parallel to the surface plane.  A Gaussian broadening of 0.01 eV is adopted.}
\label{fgr:bulk}
\end{figure}

In order to investigated how the excitonic effects are influenced by the quantum size confinement, we discuss the optical properties of bulk C$_{2}$N-\textit{h}2D, as the periodicity evolves from 2D to 3D. The optical absorption spectrum was presented in Figure~\ref{fgr:bulk} for light polarization parallel to the surface plane. There is no dark excitons in bulk C$_{2}$N. The energy of first bright exciton is 2.02 eV, with a binding energy of 0.04 eV. This value is in agreement with the experimentally optical gap (1.96 eV) for layered crystal containing more than 10 layers. It should be pointed out that the indirect interband transitions of electrons were not concerned in the present study,  which may slightly red shift the absorption peak. The $I^{1}$ transition from VBM to CBM takes place near the \textbf{k} point  (0 0 1/3) along $\Gamma$ to A, which is consistent with the minimum  QP band direct gap point just discussed above. The second excitonic peak is located at 2.14 eV with a binding energy of 0.02 eV. Such small binding energies indicate that excitons in bulk C$_{2}$N are typical Mott-Wannier excitons.

A power law fitting of the same form  is applied for excitonic energies of $I^{1}$ (optical gap). The evolution of optical gaps follows a $1/N^{0.90}$ power law, which means a faster reduction than QP band gap. An empirical formula 
\begin{equation}
E_{b}=0.21*E_{gap}+0.4
\label{eq1}
\end{equation}
was proposed by Ref. \cite{zzy-prl}, which shows a linear law between $E_{b}$ and QP band gap $E_{gap}$ for 2D materials. For monolayer our calculated $E_{b}$ is in good agreement with Equation~\ref{eq1}, while $E_{b}$ of bilayer and trilayer are at least 0.3 eV smaller than those predicted according to the empirical formula. The failure of Eq.~1 is mainly attributed to their assumption that macroscopic dielectric constant $\epsilon$ of layered structures equals to 1. However, owing to the enhanced screening effects in multi-layered strucutres, $\epsilon$ varies significantly with the increasing thickness from quasi-2D to 3D.\cite{quasi-2d-e} Thus the relationship between $E_{gap}$ and $E_{b}$ for multilayer systems is no longer simple linear law.

\section{Conclusion}
In summary, we investigated electronic and optical properties of monolayer, bilayer and bulk C$_{2}$N-\textit{h}2D by using \textit{GW}+BSE methods. Large quasiparticle band gap corrections to LDA were found in all the calculated structures. By comparing to the results of HSE06 functional, it shows a deeply dependence of coulomb correlation interactions on the number of layers, while exchange interactions are almost unchanged. In layered structures, strong excitonic effects play a crucial role in optical properties, with a significant large binding energy assigned to bound excitons. As the structure evolves from monolayer to bulk, the leading role of excitonic optical absorptions becomes weaker. Different to common 2D materials, C$_{2}$N-\textit{h}2D exhibits comparable excitonic effects in both directions for light polarization parallel and perpendicular to the layer plane. Excitons in bulk turn to Mott-Wannier excitons with small binding energues. The optical gap of bulk C$_{2}$N is in agreement with the result by the experiment. As a result of quantum size confinement, quasiparticle band gaps and optical gaps reduce with the increase of layer number ($N$), following $1/N^{0.69}$ and $1/N^{0.90}$ power law respectively. Absorptions in region of visible light are found for all the layerd and bulk C$_{2}$N. We suggest layered C$_{2}$N-\textit{h}2D may have extraordinary electronic and optical properties, especially for visible light catalysis.

\begin{acknowledgement}

This work is partially supported by the National Key Basic Research Program (2011CB921404), the NSFC (21421063, 91021004, 21233007), by CAS (XDB01020300), and by USTCSCC, SCCAS, Tianjin, and Shanghai Supercomputer Centers.

\end{acknowledgement}

%\begin{suppinfo}

%\end{suppinfo}

%\bibliography{achemso-demo}

\footnotesize{
\bibliography{c2n_jpcc}
}

%\[
%\includegraphics[width=0.5\textwidth]{TOC.pdf}
%\]
%\centerline{TOC graphic}

\end{document}